\documentclass[11pt,psfig,emulateapj,apj]{emulateapj}

\slugcomment{Not to appear in Nonlearned J., 45.}

\shorttitle{Topology of Matter and `Galaxy' Distributions}
\shortauthors{Park et al.}

\begin{document}


\title{Effects of Gravitational Evolution, Biasing, and Redshift \\
Space Distortion on Topology}

\author{Changbom Park, Juhan Kim}
\affil{Korea Institute for Advanced Study, Dongdaemun-gu, Seoul 207-43, Korea}

\and
\author{J. Richard Gott III}
\affil{Department of Astrophysical Sciences, Peyton Hall, Princeton University,
Princeton, NJ 08544, USA}




\begin{abstract}
We have studied the dependence of topology of large scale structure on
tracer, gravitational evolution, redshift space distortion, and cosmology.
A series of large N-body simulations of the $\Lambda$CDM and SCDM models
that have evolved 1.1 or 8.6 billion particles, are used in the study. 
Evolution of the genus statistic, used as a topology measure, from 
redshift 8 to 0 is accurately calculated over a wide range of smoothing scales
using the simulations.
The tracers of large scale structure considered are the CDM matter,
biased peaks in the initial density field, dark halos, and `galaxies'
populating the dark halos in accordance with a Halo Occupation Distribution
(HOD) model.

We have found that the effects of biasing, gravitational evolution, and 
initial conditions on topology of large scale structure are all comparable.
The redshift space distortion effects are relatively small down to 
about 5 $h^{-1}$Mpc for all tracers except for the high threshold part 
of the genus curve.
The gravitational effects are found to be well-modeled by the analytic
perturbation theory when the CDM matter distribution is considered. 
But the direction of gravitational evolution of topology can be even reversed
for different tracers. For example, the shift parameter of the genus curve
evolves in opposite directions for matter and HOD `galaxies' at large scales.
At small scales there are interesting 
deviations of the genus curve of dark halos and `galaxies' from that of matter 
in our initially Gaussian simulations. 
The deviations should be understood as due to combined effects of
gravitational evolution and biasing.
This fact gives us an important opportunity: topology of large scale 
structure can be used as a strong constraint on galaxy formation mechanisms.
At scales larger than 20 $h^{-1}$Mpc all the above effects gradually decrease.
With a good knowledge of the effects of non-linear gravitational evolution 
and galaxy biasing on topology one can also constrain
the Gaussian random phase initial conditions hypothesis to high accuracy.
\end{abstract}





\section{Introduction}

Topology analysis has long been applied to galaxy redshift survey data to test 
one of the major predictions of simple inflationary scenarios: that the 
primordial density fluctuations are Gaussian random phase 
(Gott, Melott, \& Dickinson 1986; Gott et al. 1989; Vogeley et al. 1994). 
To recover the primordial density 
fluctuations it is necessary to smooth the observed sample over large 
enough scales to reach the linear regime. Since redshift surveys have 
not been very deep in the past, one was left with a small dynamic range 
for topology study after big smoothings, and it has been hard to draw firm 
conclusions on the nature of the initial density field. 
As ambitious new redshift surveys like the 2dF and the Sloan Digital Sky 
Surveys(http://sdss.org) are being completed, it is now hoped that the 
topology analysis can become one of the precision measures for cosmology. 
These surveys are not just large in angle and deep in depth, but are 
also dense. This  means that the new survey 
samples can be used to study the primordial density field at large scales, 
and also to explore small-scale phenomena like the formation of galaxies.

Since the main purpose of studying topology has been to discover the 
Gaussianity of the linear density field, there has been relatively little 
work on topology at small non-linear scales. One work on small scale 
topology is Springel et al. (1998) who applied the analysis to their 
N-body simulations and the 1.2-Jy IRAS galaxy sample. However, without 
quantitatively studying the complete scale- and cosmology dependences of 
topology, as measured by the genus, they have incorrectly concluded that 
the shape of the genus curve is that of a random phase field far into 
the non-linear regime, and that the amplitude drop of the genus curve 
due to the phase correlation is the only sensitive measure of non-linearity. 
In this paper we will show that topology analysis becomes useful 
when a sample is studied over a wide range of smoothing scale and 
when quantitative measures of the deviation of the genus curve from the 
Gaussian one are used. In particular, we will present our finding that 
topology of underdense regions at small scales is sensitive to the 
structure formation mechanism. This fact opens the possibility that 
galaxy formation can be better understood by studying the small scale 
topology of galaxy distribution.

Previous analyses of observational samples for topology study have 
generally used the dark matter distribution for comparison. 
The similarity of the observed genus curves to those obtained from 
simulated matter distribution has been investigated for various 
cosmogonies. This may be fine when the uncertainties in genus 
measured from observations are much larger than the effects of 
biasing and redshift space distortion. With the unprecedented 
accuracy provided by observational data like the SDSS sample, 
however, these effects have to be accurately accounted for. 
In this paper we will study the gravitational evolution of the genus 
of peaks in the initial density field and dark halos as well as dark 
matter to see the dependence of topology on these tracers. Redshift 
space distortion effects are also studied. They all have non-negligible, 
important effects on the genus curve.

Our work is numerical since the main purposes of our work is to explore the strong 
non-linear effects of gravitational evolution through topology analysis 
where current analytic approaches are inadequate. Study of the tracer dependence of, 
and biasing effects on, topology also requires numerical simulations. 
To check for the important question of whether the topology is consistent with 
Gaussian random phase initial conditions to high accuracy in the new large observational
data sets given that the effects of non-linear evolution and biasing are detectable
in such surveys, the best method would seem to be direct comparison with large
cosmological simulations with Gaussian random phase initial conditions where 
non-linear evolution and biasing are properly modeled. If agreement is obtained,
this supports the Gaussian random phase initial conditions hypothesis to high
accuracy.

Introduction to the genus and its related statistics is given in sections 2 and 3.
Our N-body simulations used in this study are described in section 4. Evolution of
the genus curve is studied in section 5 for matter and `galaxy' distributions
in the $\Lambda$CDM and SCDM cosmogonies. Conclusions follow in section 6.


\section{GENUS}

To measure topology of a tracer distribution we calculate the genus of iso-density 
surfaces as a function of density threshold level. The genus is defined as
\begin{eqnarray}
\nonumber
G &=& {\rm Number~of~holes~in~contour~surfaces}\\
&-&{\rm Number~of~isolated~regions}
\end{eqnarray}
in the isodensity surfaces at a given threshold level. In this definition, a sphere 
has genus of $-1$, and a torus has genus 0 (Gott, Melott \& Dickinson 1986). 
The Gauss-Bonnet theorem connects the global topology with an integral of local 
curvature of the surface $S$, i.e.
\begin{equation} G = -{1 \over {4 \pi}} \int_S \kappa dA, \end{equation}
where $\kappa$ is the local Gaussian curvature. The genus is equal to $-1/2$ times 
the Euler characteristic, or the fourth Minkowski Functional in 3-dimensions (Mecke,
Buchert, \& Wagner 1994; Schmalzing \& Buchert 1997).

   For a Gaussian random field an analytic formula exists for the genus-threshold 
density relation (Hamilton, Gott, \& Weinberg 1986). The genus per unit volume is 
\begin{equation}
g(\nu) = A (1-\nu^2 ) e^{-\nu^2 /2}, 
\end{equation}
where $\nu \equiv (\rho-{\bar\rho})/\sigma$ is the threshold density in unit of 
standard deviations $\sigma=\langle(\rho-{\bar\rho})^2\rangle^{1/2}$ 
from the mean. The amplitude is 
\begin{equation} 
A = {1 \over {(2 \pi)^2}} ({{\langle k^2\rangle} \over 3})^{3/2}, 
\end{equation}
where the second moment
\begin{equation}
{\langle k^2\rangle} = \int P(k) W(k) k^2 d^3 k / \int P(k) W(k) d^3 k 
\end{equation}
depends only on the shape of the power spectrum and the smoothing kernel $W$. 
Instead of using $\nu$ to identify isodensity contours we use the volume fraction 
label $\nu_f$ at which the corresponding isodensity contours enclose the fraction 
of the sample volume equal to that at the density threshold $\nu_f \sigma$ 
in the case of a Gaussian field.  In this way the genus curve becomes independent 
of the one-point probability distribution function, which can be better studied 
by different statistics. Use of $\nu_f$ also makes the genus curves less 
sensitive to details of bias (Park \& Gott 1991), which is beneficial in 
studying the primordial field.

\section{GENUS-RELATED STATISTICS}
\subsection{Shift Parameter}
  Any departure of the genus curve from the relation given by equation (3) indicates non-Gaussianity of the field. To quantify the deviation we define a set of genus-related statistics. The first statistic $\Delta\nu$ measures the shift of the middle part of the genus curve.
We compute $\Delta\nu$ defined by (Park et al. 1992)
\begin{equation}
\Delta\nu = \int_{-1}^{1} \nu G_{\rm obs}(\nu) d\nu / \int_{-1}^1 G_{\rm fit}(\nu)d\nu, 
\end{equation}
where $G_{\rm obs}(\nu)$ is the measured genus and $G_{\rm fit}(\nu)$ is the
random-phase curve best fit to the measured genus data from $\nu=-1$ to $+1$.
The best fitting amplitude $A_{\rm obs}$ of the observed genus curve is 
obtained from $G_{\rm fit}(0)$.

\subsection{Amplitude Drop}
The second statistic is the amplitude drop of the genus curve. The density field with 
phase correlation due to gravitational evolution typically has a fewer number 
of structures compared to that with the same power spectrum but with random phases. 
A derived statistic is 
\begin{equation}
R_{\rm A} = A_{\rm obs} /A_{\rm PS}, 
\end{equation}
where $A_{\rm PS}$ is the amplitude expected for a Gaussian field which has the 
power spectrum (PS) of the evolved particle/`galaxy' distribution. Existence of phase 
correlation reduces $R_{\rm A}$ below 1 (Vogeley et al. 1994; Canavezes et al. 1998).
The best fitting amplitude $A_{\rm obs}$ is measured as explained above.   
To calculate $A_{\rm PS}$ the PS of the distribution is needed. 
In this study tracers are contained in simulation cubes with periodic boundaries,
and the PS is obtained by Fourier transforming the distribution.
Extrapolations of the PS beyond the observed $k$-window necessary to calculate 
the integral for $A_{\rm PS}$ introduce only a small error in $A_{\rm PS}$  
when ${\bar d}/\sqrt{2} \le R_G \le r_{\rm max}/10$ where $R_G$ is the Gaussian smoothing length,
${\bar d}$ is the mean tracer separation, and $r_{\rm max}$ is the sample size.

\subsection{Cluster/Void Abundance Parameters}
The third statistic is the numbers of clusters and voids. The statistic $A_C$ is defined as
\begin{equation}
A_C = \int G_{\rm obs}(\nu) d\nu / \int G_{\rm fit}(\nu) d\nu,  
\end{equation}
where the integral is limited to $1.2\le \nu\le 2.2$ which is roughly centered at 
$\nu=\sqrt{3}$ which is the minimum in the best fit random phase curve where the number of
isolated clusters is greatest. 
$A_V$ is similarly defined over $-2.2\le \nu\le -1.2$ roughly centered at 
$\nu=-\sqrt{3}$, the minimum in the best fit random phase curve where 
the number of isolated voids is greatest. 
Here the best-fit Gaussian curve $G_{\rm fit}$ is obtained as explained above 
using the middle part of the genus curve.  The statistics $A_C$ and $A_V$ 
quantify the asymmetry of the genus curve and the multiplicity of clusters and voids.

\section{SIMULATIONS}

In order to study the behaviour of the genus statistic accurately 
we have made a set of 
large N-body simulations using a PM and PM+Tree codes. Our PM code uses the 
triangular-shaped cloud method for mass assignment and force interpolation 
at particle positions. The force calculation at mesh points is done by the 
four-point finite difference algorithm (Park 1990, 1997). The PM+Tree code GOTPM, 
which is a merged version of our PM code with Dubinski's tree code, is a 
parallel code which adds the tree force to the PM force at separations 
shorter than 4 times the pixel size of the mesh used in the PM part 
(Dubinski et al. 2003). For topology study of the matter distribution, 
the PM code is preferred since this code can make simulations with large 
dynamic range in mass at relatively low cost. The PM code is also 
preferred when the peaks in the initial density field are used as biased 
tracers because high-resolution initial density fields are adopted
in this code. Due to its high force resolution power the PM+Tree code should 
be the choice when dark halos are targets for topology study.
Because a single simulation cannot satisfy all of our needs we have made 
several simulations supplementing one another using the PM and PM+Tree codes. 
Parameters of our simulations are summarized in Table 1.
\begin{table}
\footnotesize
\begin{center}
\caption{Characteristics of Cosmological N-Body Simulations}
\vspace{0.3cm}
\begin{minipage}{\linewidth}
\begin{center}
\renewcommand{\thefootnote}{\thempfootnote}
\begin{tabular}{l| c c c c c c c c c c c}
\hline\hline
Name
& $\Omega_m$
& $\Omega_{\Lambda}$
& $h$
& $b$
& $N_m$
\footnote{Size of mesh on which initial conditions are defined.}
&$N_p$
&$L$
\footnote{$(h^{-1}{\rm Mpc})$}
&$z_i$
&$N_{step}$
&code
\\
\hline
LCDM1024
&$0.27$
&$0.73$
&0.71
&1.11
&$2048^3$
&$2048^3$
&$1024$
&17
&680
&PMTree
\\
LCDM5632
&$0.27$
&$0.73$
&0.71
&1.11
&$2048^3$
&$2048^3$
&$5632$
&17
&170
&PMTree
\\
LCDM410
&$0.27$
&$0.73$
&0.71
&1.11
&$2048^3$
&$1024^3$
&$409.6$
&47
&470
&PM
\\
SCDM1024
&$1.00$
&$0.00$
&0.5
&1.5
&$2048^3$
&$1024^3$
&$1024$
&23
&230
&PM
\\
SCDM410
&$1.00$
&$0.00$
&0.5
&1.5
&$2048^3$
&$1024^3$
&$409.6$
&47
&470
&PM
\\
\hline
\end{tabular}
\end{center}
\end{minipage}
\label{simresolution}
\end{center}
\end{table}

Our results are mainly derived from PM+Tree simulations 
of the $\Lambda$CDM model with the WMAP parameters, namely, the density parameters 
$\Omega_m = 0.27, \Omega_{\Lambda} = 0.73$, the Hubble constant $h = 0.71$, 
$\sigma_{{\rm m}, 8}=0.9$, and the matter is dominated by the cold dark 
matter (CDM). Here $\sigma_{{\rm m}, 8}$ is the RMS mass fluctuation in 
a sphere of radius of 8 $h^{-1}$ Mpc. 
The physical sizes of the two simulation cubes are $L=1024$ and $5632 h^{-1}$ Mpcs. 
We call them LCDM1024 and LCDM5632.
We use the transfer function of the power spectrum given by Bardeen et al. 
(1986) for LCDM1024. But we have used Eisenstein \& Hu (1998)'s power spectrum 
with $\Omega_{\rm baryon}=0.0463$ for LCDM5632 to include baryonic effects. 
The number of CDM particles evolved in both simulations is $2048^3 = 8.59$ 
billion, among the largest N-Body simulations ever made. 
A $2048^3$ mesh is used for the PM part, and the force softening length in 
the Tree part is one tenth of the mean particle separation. Therefore, the 
dynamic range in length is $2 \times 10^4$, from 0.05 to 1024 $h^{-1}$ 
Mpc while the dynamic range in mass is $8.6 \times 10^9$, from 
$9.4\times 10^9$ to $8.0\times 10^{19}$ $h^{-1}\rm M_{\odot}$ in the case
of LCDM1024, for example.
Dark halos are identified in the LCDM1024 simulation made by the PM+Tree code. 
Halo centers are found as local density peaks defined on a fine mesh.
The CDM particles belonging to virialized dark halos are searched by 
checking their binding energy to the local halo centers and tidal radii 
of subhalos with respect to adjacent bigger halos (Kim \& Park 2005). 
These identification criteria overcome many of the problems and artificial results 
obtained from existing group-finding algorithms, and let us obtain a catalog 
of self-bound physical dark halos. Only the dark halos with more than 53 
member particles or heavier than $5\times 10^{11}$ $h^{-1} \rm M_{\odot}$ are adopted. 

Even though our group finding algorithm allows us to identify many subhalos 
contained in big halos, those halos with mass much higher than $10^{13} M_{\odot}$ 
cannot be considered to be a single galaxy. On the other hand, every halo 
identified in the N-body simulation might not contain a galaxy. Therefore, 
even though dark halos might be better tracers of the observed galaxies 
than matter, they are still biased tracers.
We address this problem by introducing the Halo Occupation Distribution (HOD)
prescription to locate `galaxies' within our dark halos.
We adopt the following recipe of Zehavi et al. (2004).
A central galaxy is assigned to a halo if its mass $M$ exceeds $M_{\rm min}$,
and the central galaxy can have satellites. The mean number of satellite
galaxies is given by a power-law $\left< N_{\rm sat} \right> = (M/M_1)^{\alpha}$, 
and they have a Poisson distribution. 
We use, as a toy model, the HOD parameters of
${\rm log} M_{\rm min} =11.76$, ${\rm log} M_1 =13.15$, and $\alpha=1.13$
given by Table 3 of Zehavi et al. corresponding to galaxies with an absolute 
$r$ magnitude cut of $M_r = -19.5$ in the Sloan Digital Sky Survey. 
These `galaxy' tracers are used in subsection 5.3.

As an another galaxy formation model we adopt the biased galaxy formation 
scenario where galaxies form at peaks in the initial density field 
(Bardeen et al. 1986). We use PM simulations of $1024^3$ CDM particles 
in a $2048^3$ mesh spanning $409.6 h^{-1}$ Mpc along a side. In the 
$\Lambda$CDM model a set of peak bias parameters, ${\nu_{\rm th} = 0.05}$ 
and $R_s = 0.763 h^{-1}$ Mpc, gives us peak particles which cluster 
like bright galaxies at the present epoch, where $\nu_{\rm th}$ 
is the threshold density level for the peaks, and $R_s$ is the 
Gaussian smoothing length corresponding to the galaxy scale. 
In the case of the SCDM model, the peak 
parameters adopted are ${\nu_{\rm th} = 1.7}$ and $R_s = 0.585 h^{-1}$ 
Mpc for $\sigma_{{\rm mass}, 8} = 0.6$. These peak particles have been 
shown to follow the centers of collapsed dark halos, but are too clustered 
in high density regions since no merging process is included 
(Park \& Gott 1991).

\section{RESULTS}
We have measured the genus statistics in the four dimensional space of smoothing 
scale, redshift, tracers, and cosmogony. We use Gaussian smoothing and the 
definition of the smoothing length $R_G$ is standard ($\sqrt{2}$ times 
larger than the `e-folding' smoothing adopted by Gott et al. (1989)). We 
always maintain the condition that the Gaussian smoothing length $R_G$ must 
be greater than or equal to the mean particle/halo separations. Topology is 
explored at 14 smoothing scales from $R_G = 1.5 h^{-1}$ Mpc to 30 $h^{-1}$ 
Mpc in the case of the matter distribution. Our large and high resolution 
simulations enable us to obtain very accurate genus values. The number of 
resolution elements in the simulation box is over 2500 
($=L^3 /(2\pi)^{3/2} R_G^3$) at the largest smoothing scale $R_G = 30 h^{-1}$ 
Mpc, and exceeds $2\times 10^7$ at $R_G = 1.5 h^{-1}$ Mpc.
The genus is also calculated at 7 epochs with redshifts $z$ from 0 to 8, 
but we show results only at $z=0, 1, 2, 5,$ and 8 to avoid over-crowding 
of points in figures.

\subsection{$\Lambda$CDM Dark Matter}
The mean particle separation of the LCDM1024 simulation is 0.5 $h^{-1}$ Mpc, 
and we explore the topology of the dark matter distribution 
from $R_G = 1.5 h^{-1}$ to 30 $h^{-1}$ Mpc. The size of density 
array is $2048^3$, so the smoothing length is at least three times 
larger than the pixel size (thus avoiding discreteness effects). 
The results for the genus-related statistics are shown in Figure 1 as filled
and open circles.
\begin{figure}
\begin{minipage}[t]{7cm}
\epsfxsize=8cm
\epsfysize=8cm
\epsffile{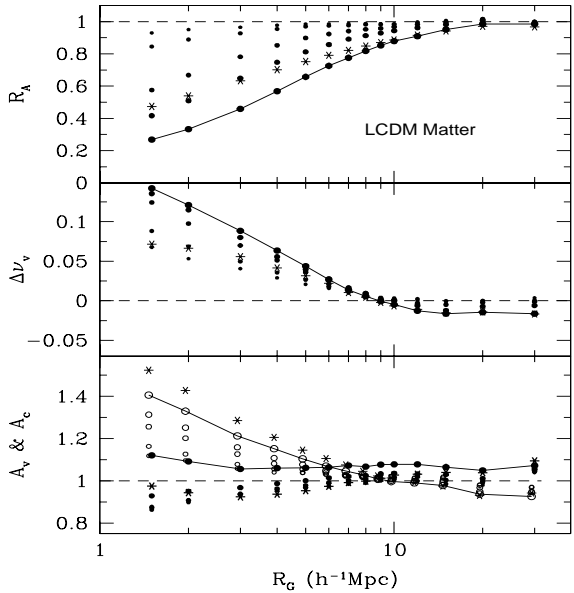}
\end{minipage}
\caption{The genus-related statistics for dark matter measured from 
the LCDM1024 simulation. Dots with different sizes represent the statistics
at redshifts $z=0, 1, 2, 5$, and 8. The largest dots connected by lines are
for $z=0$, and the smallest are for $z=8$. In the bottom panel filled circles
are $A_C$'s, and open circles are $A_V$'s which are slightly shifted to avoid
confusion. Stars are the statistics measured in redshift space. In the bottom
panel bold stars are $A_C$'s, and thin stars are $A_V$'s.
\label{fig1}}
\end{figure}
Dots with different sizes represents the statistics at different 
redshifts from $z=0$ (largest circles connected by solid lines), 
1, 2, 5 to 8 (smallest). The dark 
matter in the $\Lambda$CDM universe clearly shows an intriguing development 
of small-scale topology 
in the cases of $\Delta\nu$ and $A_V$ statistics
in a direction opposite to that at large scales. 
The shift parameter $\Delta \nu$ rapidly goes to positive 
values at $R_G <9 h^{-1}$ Mpc. But at scales $R_G > 9 h^{-1}$ Mpc it 
slowly moves in the negative direction and continues to be negative 
even at 30 $h^{-1}$ Mpc scale. All these genus curves have a spongelike
topology ($G >0$) at the median density contour, so we would not call 
the negative and positive shifts of the central part of the genus curve 
as in the direction of 'meat-ball' and 'bubble' topologies, respectively, 
since their correspondence does not seem to be so simple.
The transition scale of $\Delta\nu$ from positive to negative shift
agrees with the analytic prediction of subsection 5.6.

\begin{figure}
\begin{minipage}[t]{7cm}
\epsfxsize=8cm
\epsfysize=8cm
\epsffile{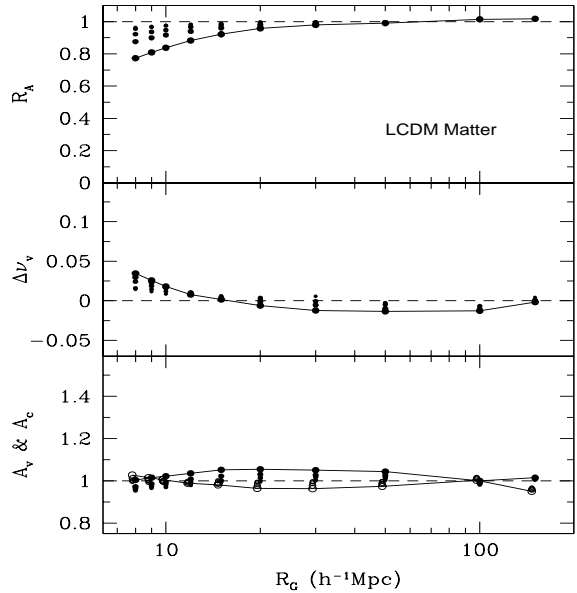}
\end{minipage}
\caption{Same as Figure 1, but statistics are calculated from 
the LCDM5632 simulation. Note that the data at $z=8$ and 5 are almost 
overlapping. Only measurements in real space are plotted.
\label{fig2}}
\end{figure}
In Figure 2 the genus-related statistics calculated from the LCDM5632
simulation are plotted to show their behaviour at larger scales.
Note that the power spectrum of LCDM5632 is slightly different from
that of LCDM1024, thus making the transition scale of $\Delta\nu$
occur at different $R_G$. It can be seen that non-Gaussianity is
visible up to the scale $\sim 100 h^{-1}$Mpc. But using the analytic
estimation of the non-Gaussianity effects of gravitational evolution
in this weakly non-linear scales given in subsection 5.6, one could
draw conclusions on the Gaussianity of the initial density field
from data with smoothing lengths much smaller than $100 h^{-1}$Mpc 
if the matter distribution can be used.

The parameter $A_V$ for voids (open circles in the bottom panels of Figure 1
and 2) shows a behaviour similar to the shift parameter. 
At scales below 10 $h^{-1}$ Mpc voids are broken 
into many pieces and the number of voids exceeds the Gaussian value 
predicted by the middle part of the genus curve (see Figure 1). But at 
larger scales voids percolate and the number of voids is slightly less 
than the Gaussian value (see Figure 2). On the other hand, the parameter 
$A_C$ for clusters (filled circles in the bottom panel of Figure 1 and 2) 
behaves differently. It tends to increases at all 
scales at the present epoch. The speed with which new high density 
clumps stick up above the threshold and appear in the sample is faster 
than that of merging and disappearance of existing clumps. 
But at very small scales ($R_G \le 2 h^{-1}$ Mpc) the merging speed wins 
over the sticking-up speed at high redshifts ($z \ge 5$) and subsequently
the situation reverses. At large scales, the sticking-up of clusters 
and merging of voids are the dominant processes (Figure 2). 
It should be noted that gravitational evolution only makes the both 
parameters $A_V$ and $A_C$ increase over 1 at scales below 10 $h^{-1}$ Mpc 
at the present epoch. Therefore, the observed value of the void abundance 
parameter $A_V$ which is less than 1 (Park et al. 2005) must be due to 
strong biasing effects, 
which tend to make small voids merge with one another. This can be achieved 
if galaxies do not form in weak walls and filaments of large scale structure 
dividing small voids.

The amplitude drop parameter shown in the top panel of Figure 1 evolves
very rapidly at scales below about 10 $h^{-1}$Mpc as the gravitational
evolution proceeds. And its scale dependence is again determined by the 
characteristic shape of the power spectrum of the $\Lambda$CDM model.

Also shown in the figures are the genus-related statistics measured in
redshift space at $z=0$ (stars in all figures). We adopt the far-field approximation
to the redshift distortion of the density field. Displacement of particles
is thus assumed to occur only along the direction of a coordinate axis.
We average three genus curves measured from redshift space density fields
each of which is
perturbed in the direction of each coordinate axis using the periodic
boundary condition. Genus and the power spectrum of density field in redshift space
are measured to calculate the genus-related statistics.
The amplitude drop parameter $R_A$ increases in redshift space because 
the amplitude of the random-phase genus curve ($A_{\rm PS}$ in Equation (7)) 
drops more than that of non-linear genus curve ($A_{\rm obs}$) 
due to the randomization effects of small-scale peculiar velocity.
Effects of redshift space distortions on the genus curve have been
previously studied by Melott, Weinberg, \& Gott (1988).
They have reported that the effects are negligible when a smoothing
larger than the correlation length is applied to the density field.
And the linear regime calculation of Matsubara (1996) has supported 
their claim.  Subsequently, redshift space distortion effects have been
generally ignored in the genus analyses.
This is qualitatively true for the matter field of the $\Lambda$CDM
model where the correlation length is about 5.5 $h^{-1}$Mpc at $z=0$
and smoothing lengths larger than this are studied.
However, as can be seen in our figures, there are non-negligible systematic 
effects of redshift space distortion on the genus related statistics
at the small smoothing scales we are interested in.
An example is the $A_C$ parameter (thick stars in the bottom panel of 
Figure 1). Being a measure of the topology
of high density regions, $A_C$ is most sensitive to the peculiar velocity field.

\subsection{$\Lambda$CDM Biased Peak Particles}
We use the peaks in the initial Gaussian density field
as the biased `galaxy' particles.
They are defined by the threshold level $\nu_{\rm th}$ and the galaxy scale 
$R_s$ (Park \& Gott 1991), which are chosen so that 
the resulting biased particles 
approximate the distribution of galaxies at the present epoch. When we require 
them to have $\sigma_{{\rm pk},8} = 1$ and to have a correlation function 
close to the observed galaxy CF at $z=0$, we find a set, $\nu_{\rm th} = 0.05$ 
and $R_s = 0.763 h^{-1}$ Mpc in our $\Lambda$CDM universe with 
$\sigma_{\rm m, 8} = 0.9$. The number density of the peaks is much lower than 
that of CDM particles, and the genus is measured only at scales $R_G \ge 5 h^{-1}$ Mpc.
The simulation used for identifying the biased peak particles is 
the $\Lambda$CDM model evolved by the PM code in a $2048^3$ mesh 
spanning $409.6 h^{-1}$ Mpc along a side. Its small pixel size 
($0.2 h^{-1}$ Mpc) allows us to look for density peaks at the galaxy scale $R_s$.
The genus-related statistics of the biased peak particles are shown in Figure 3. 
Even though our biased density peaks collapse at a specific epoch and can be
considered as galaxies after then, we plot the statistics of the biased particles
at all epochs after $z=8$ for theoretical interest.
\begin{figure}
\begin{minipage}[t]{7cm}
\epsfxsize=8cm
\epsfysize=8cm
\epsffile{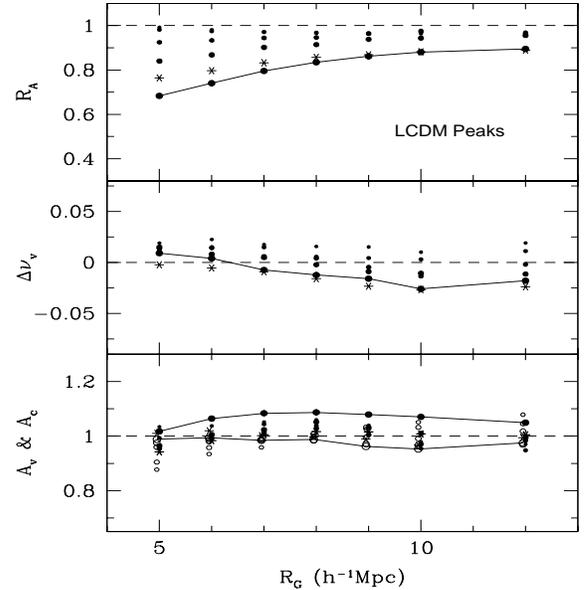}
\end{minipage}
\caption{Genus-related statistics for the biased peaks
calculated from the $\Lambda$CDM410 simulation.
Notations are the same as Figure 1.
\label{fig3}}
\end{figure}

They behave essentially in the same way as those for cold dark matter, 
but there are some changes. Similarly to before, the peak particles 
develop a negative shift $\Delta \nu$ at scales larger than $6 h^{-1}$Mpc, 
and a positive shift at smaller scales. And the cluster abundance 
parameter $A_C$ increases at all scales above $5 h^{-1}$ Mpc. 
However, the void abundance parameter $A_V$ (open circles in Figure 3) shows more
changes. Namely, at scales smaller than about 7 $h^{-1}$Mpc the biasing
prescription makes small voids percolate and $A_V$ become less than 1
at high redshifts.  As noted 
above, gravitational evolution only makes $A_V$ increase above 1 at scales less 
than 10 $h^{-1}$Mpc, contrary to observations. Here we have a specific example 
of a galaxy formation mechanism which allows existing voids to percolate strongly 
enough to overcome the nonlinear gravitational effect of opening of new voids
and makes $A_V$ less than 1.

The stars in Figure 3 are the statistics measured in redshift space at $z=0$. 
The effect significantly decreases $A_C$ (bold stars in the bottom panel of Figure
3) at all scales explored as dense clusters are spread along fingers-of-god. 
It can be noted that the void abundance parameter is not much
affected by the redshift space distortion because empty regions are much more 
stable to the distortion compared to dense clumps. This observation is also
true for other tracers used in this work. Given its insensitivity 
to redshift space distortion, $A_V$ is a crucial statistic telling the story 
of galaxy formation in a less distorted way.

\subsection{$\Lambda$CDM Dark Halos}
The biased particles studied above are the density peaks in the Gaussian initial 
density field. This structure formation scheme assumes that galaxies form 
at the maxima in the primordial density fluctuations. It can be shown 
that these actually correspond well to collapsed dark halos at the final 
epoch of simulations (Park \& Gott 1991). However, it would be good to check the 
idea and to re-compute the same statistics by using physically bound 
dark halos identified in high resolution simulations. 
Another important difference between the biased peak particles and dark halos
is that formation of tracers is naturally incorporated in the genus analysis
in the case of dark halos.
Since we need high resolution for halo formation and identification and large 
volume for high statistical significance, the LCDM1024 simulation is used.
The minimum halo mass identified is
$5 \times 10^{11} M_{\odot}$ (53 particles). 
With the dark halo samples obtained from this
simulation, topology of the dark halo distribution is studied at 
scales from 5 to 30 $ h^{-1}$ Mpc at $z=0$. At higher redshifts
the minimum smoothing length used becomes larger as the dark halos collapsed
by then become rarer.

The genus-related statistics measured for the dark halos in the $\Lambda$CDM 
model are shown in Figure 4. Topology of $\Lambda$CDM dark halos is
very different from those of underlying dark matter or biased peak particles.
The shift parameter $\Delta \nu$ starts from large negative values
at high redshifts when the only densest regions can collapse.
Consistent with this behaviour the $A_C$ parameter is greater than 1,
and $A_V$ is lesser than 1 at high redshifts.
At lower redshifts, the topology of the dark halo distribution evolves strongly.
$\Delta\nu$ becomes positive and $A_C$ becomes less than 1. This is
opposite to what we have seen in the case of CDM matter
and biased peak particle distributions. $A_V$ still
stays below 1 but has increased from its high redshift value, which is consistent
with the case of biased peak particles.
$A_C$ of dark halos rapidly decreases at low 
redshift while that of peak particles keeps increasing. This is because 
the dark subhalos merge to form one huge halo at centers of clusters of 
halos, and thus decreasing the number of subhalos or reducing the richness 
of clusters as non-linear evolution proceeds at small scales. This 
merging process within clusters does not exist for peak particles because 
peak particles never lose their identity and the clusters of peak particles 
get richer as they become more compact. The reality might be somewhere 
between these two extremes. 
Comparison of Figure 4 with previous ones demonstrates 
sensitivity of the topology to the galaxy formation mechanism.
Since galaxies show different topology depending on their internal physical
properties, topology analysis can be used to discriminate among different
galaxy formation mechanisms for different species of galaxies.
\begin{figure}
\begin{minipage}[t]{7cm}
\epsfxsize=8cm
\epsfysize=8cm
\epsffile{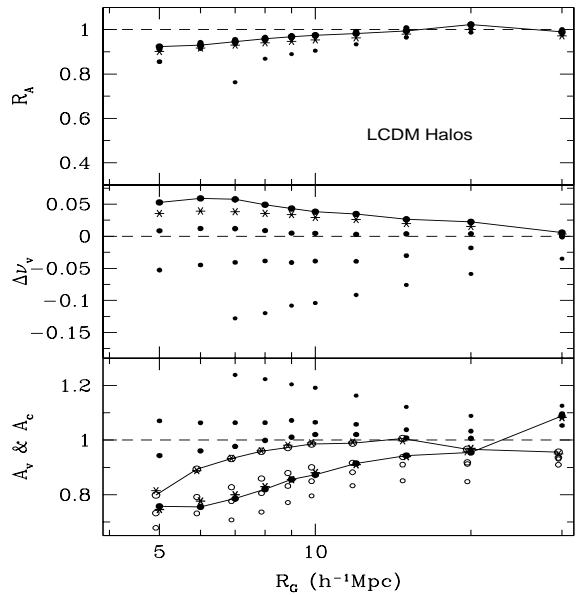}
\end{minipage}
\caption{Genus-related statistics for the dark halos
calculated from the LCDM1024 simulation. 
Notations are the same as Figure 1.
Since the number density of dark halos at $z=8$ and 5 is too low, we show 
the results at $z=0, 1, 2,$ and 3.5. And at $z=3.5$ (smallest points) genus 
is measured at $R_G \ge 7 h^{-1}$ Mpc.
\label{fig4}}
\end{figure}

Hikage, Taruya \& Suto (2003) have calculated the biasing and genus of 
dark matter halos as a function of halo mass and smoothing scale by 
using the Hubble Volume Simulation. They have found that the halo 
biasing effects on the genus are comparable to the non-Gaussianity 
due to the non-linear gravitational evolution. This is consistent with
our results that changes of the genus-related statistics due to
gravitational evolution at $R_G= 30 h^{-1}$Mpc scale is of the order of 
that due to tracer difference. 
Hikage et al. (2003) have studied topology of halo distribution 
at $R_G \ge 30 h^{-1}$ Mpc.
However, at $R_G=30 h^{-1}$ Mpc their 
results are thought to be affected by discreteness effects because the 
mean separation of their halo subsets is only about 44 $h^{-1}$ Mpc.
In topology analysis of a set of points, the smoothing length should be larger 
than the mean tracer separation to avoid systematic discreteness 
effects, or some correction should be made to remove the effects (Park et al. 2005).
The large pixel size of $23 h^{-1}$ Mpc used in their analysis may have 
also caused some discreteness effects at that smoothing scale.
In general one wants the smoothing length $R_G$ to be at least $1.77$ 
times the pixel size to avoid discreteness effects (eg. Hamilton, 
Gott, \& Weinberg 1986). 
To identify halos they have used the friends-of-friends algorithm, 
which is known to miss subhalos within big halos and often produces 
unphysical halos connected by narrow chains of particles. 
Therefore, it is not easy to directly compare their results with those of 
the current work in detail. But our results qualitatively agree with their 
conclusion that non-Gaussianity induced by halo biasing is comparable 
to that by the non-linear gravitational evolution.


\subsection{HOD Galaxies}
A halo identified in the N-body simulation might not contain a galaxy, or can
host more than one galaxies. Therefore, 
even though dark halos might be a better tracer of the observed galaxies 
than matter, they may not be representing galaxy distribution well enough
to allow quantitative comparisons.
We address this problem by introducing the Halo Occupation Distribution (HOD)
prescription to locate `galaxies' within our dark halos.
We adopt the following recipe of Zehavi et al. (2004).
A central galaxy is assigned to a halo if its mass $M$ exceeds $M_{\rm min}$,
and the central galaxy can have satellites. The mean number of satellite
galaxies is given by a power-law $\left< N_{\rm sat} \right> = (M/M_1)^{\alpha}$, 
and they have a Poisson distribution. 
The satellites are randomly located at distances of $0.3 h^{-1}{\rm Mpc}$
from the central galaxy. This does not change our results because the minimum
smoothing scale adopted for analysis of the HOD galaxies is $5 h^{-1}{\rm Mpc}$.
We use, as a toy model, the HOD parameters of
${\rm log} M_{\rm min} =11.76$, ${\rm log} M_1 =13.15$, and $\alpha=1.13$
given by Table 3 of Zehavi et al. corresponding to galaxies with an absolute 
$r$ magnitude cut of $M_r = -19.5$ in the Sloan Digital Sky Survey. 

In Figure 5 we show the genus-related statistics calculated from 
distribution of `galaxies' assigned within dark halos of the LCDM1024
simulation according to the HOD prescription. 
The correlation length of these model galaxies is $5.6 h^{-1}$Mpc,
similar to that of bright optical galaxies.
Only the results at $z=0$ are shown. Comparing
the statistics with those of dark halos at $z=0$, we find $\Delta\nu$
becomes less positive, $A_C$ is increased, and $A_V$ is decreased.
These changes the topology of our HOD `galaxies' to be more compatible with
that observed for SDSS galaxies (Park et al. 2005).
Topology depends on the type of observed galaxies, but one phenomenon consistently
found in observations is that the $A_V$ parameter is less than 1. And
This behaviour is seen in the distribution of our HOD `galaxies.
However, the observed $A_C$ parameter is typically greater than $A_V$, which does
not quite agree with the HOD `galaxies'.
The HOD method seems to be promising in explaining not only the one-point and two-point
functions of galaxy distribution, but also functions of high-order moments
like genus. However, more detailed comparisons have to be made to confirm it.
\begin{figure}
\begin{minipage}[t]{7cm}
\epsfxsize=8cm
\epsfysize=8cm
\epsffile{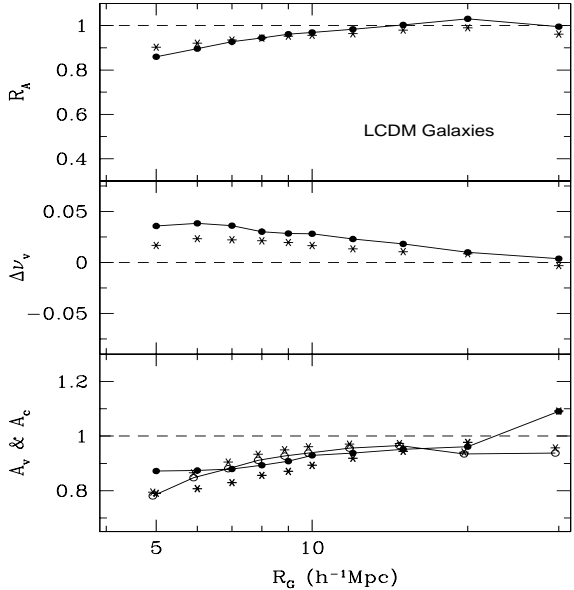}
\end{minipage}
\caption{Genus-related statistics for the `galaxies' 
derived from the dark halos of the LCDM1024 simulation.
Notations are the same as Figure 1.
Only the results at $z=0$ are shown.
\label{fig5}}
\end{figure}

Benson et al. (2001) has used a semi-analytic model of galaxy formation to 
obtain the distribution of model galaxies, and to measure the genus curves
at smoothing lengths of 4.24 and 5.66 $h^{-1}$Mpc at redshifts $z=0$ and 1. 
But they have not been able to detect the difference between the genus curves 
for the dark matter and the model galaxies 
due to sparseness and smallness of their simulation data.

\subsection{SCDM Case}
As a fiducial model we consider the SCDM model to study the dependence of 
topology on cosmology. In this model $\Omega_m = 1$ and $h=0.5,$ and we adopt
a bias factor $b=1.5$ and Bardeen et al. (1996)'s fitting formula for the power
spectrum.  Figure 6 shows that the matter distribution of the 
SCDM universe has qualitatively the same topology as that of the 
$\Lambda$CDM universe. Only the characteristic scales are moved down in
the cases of $\Delta\nu$, and $A_V$. 
The switching from negative to positive shift occurs at $\sim 4 h^{-1}$ Mpc, 
and the switching from sub-Gaussian to super-Gaussian amplitude for the 
void part of the genus curve occurs at a similar scale. This change 
in the characteristic scales is the direct consequence of the change 
in the shape of power spectrum.
Figure 7 shows the statistics for the biased peak particles found in the 
SCDM simulation. For $\Delta\nu$ a negative shift prevails at all scales, and
the redshift space distortion effects make little change in $\Delta\nu$.
Gravitational evolution 
makes the $A_C$ parameter increase above 1 as in the $\Lambda$CDM case 
while the redshift space distortion effects make it decrease. The 
void abundance parameter $A_V$ is significantly less 1 due to the 
biasing effects at small scales, and is hardly affected by the redshift 
space distortion as in the $\Lambda$CDM case.
\begin{figure}
\begin{minipage}[t]{7cm}
\epsfxsize=8cm
\epsfysize=8cm
\epsffile{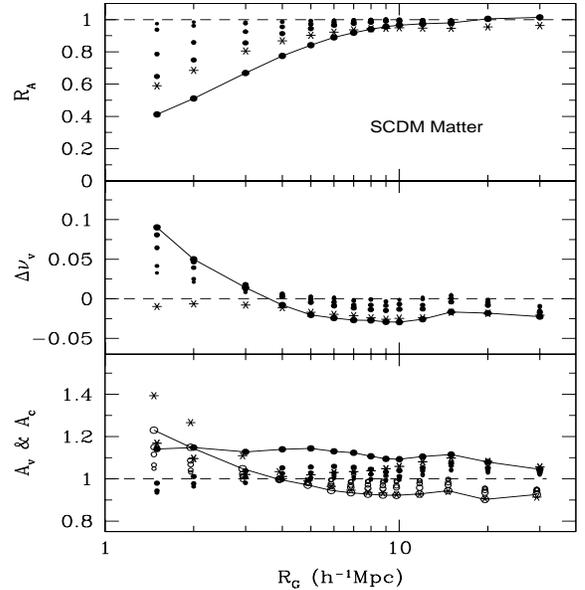}
\end{minipage}
\caption{Genus-related statistics for dark matter measured
from the SCDM1024 simulation.
Notations are the same as Figure 1.
\label{fig6}}
\end{figure}
\begin{figure}
\begin{minipage}[t]{7cm}
\epsfxsize=8cm
\epsfysize=8cm
\epsffile{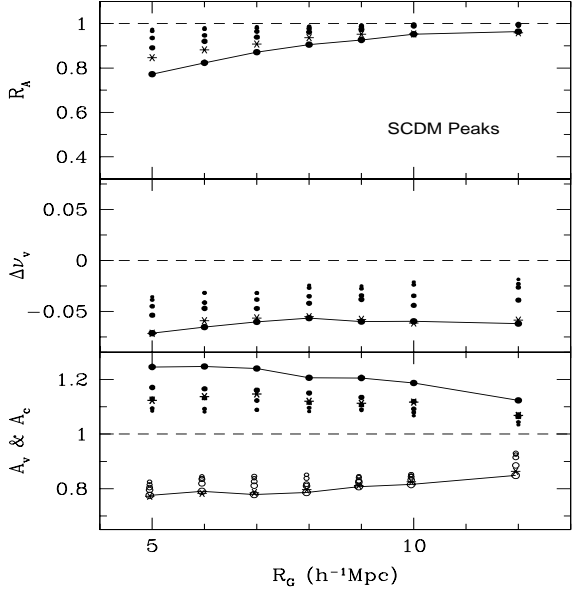}
\end{minipage}
\caption{Genus-related statistics for the biased peaks measured from
in the SCDM410 simulation.
Notations are the same as Figure 1.
\label{fig7}}
\end{figure}

\subsection{Comparison with Perturbation Theory}

Matsubara (1994) has obtained a formula for the genus curve modified due to 
gravitational evolution in the weakly non-linear regime by perturbatively 
expanding the statistic in 
$\sigma_0 = (\delta \rho/\rho)_{\rm rms}$,
the RMS fluctuation of an over-density field smoothed over a given scale. 
The non-linear correction in the genus curve to the first order in $\sigma_0$
is an odd function of $\nu$ and, therefore, causes a shift and an asymmetry 
between the high and low density regions. 
It is hoped that the gravitational evolution effects on genus in the weakly 
non-linear regime can be modeled by this analytic theory.
When the volume threshold level is used, the three-dimensional genus expanded
to the first order in $\sigma_0$ is given by (Matsubara 2003)
\begin{eqnarray}
\nonumber
G(\nu_f) &=& {1\over {(2\pi)^2}}
({\sigma_1 \over {\sqrt{3}\sigma_0}})^3 e^{-\nu_f^2 /2} (1-\nu_f^2 \\
\nonumber
&&-[(S^{(1)}-S^{(0)})(\nu_f^3-3\nu_f) \\
&& + (S^{(2)}-S^{(0)}) \nu_f]\sigma_0 ), 
\end{eqnarray}
where the variance parameters are
\begin{equation}
\sigma_j^2 (R) = \int {{k^2 dk}\over{2\pi^2}} k^{2j} P(k) W^2 (kR), 
\end{equation}
and the window function is a Gaussian $W(kR) = {\rm exp}(-k^2 R^2/2)$.
$S^{(i)}$ are the skewness parameters defined as
\begin{eqnarray}
\nonumber
 S^{(0)} &=& \left< \delta^3\right> /\sigma_0^4\\
&=&{1\over \sigma_0^4}\int{{d^3k_1}\over{(2\pi)^3}}
{{d^3k_2}\over{(2\pi)^3}}B(k_1,k_2,k_{12}), 
\end{eqnarray}
\begin{eqnarray}
\nonumber
 S^{(1)} &=& -{3 \over 4}\left< \delta^2\nabla^2\delta\right> /\sigma_0^2 
\sigma_1^2 \\
&=&{3\over {4\sigma_0^2\sigma_1^2}}\int{{d^3k_1}\over{(2\pi)^3}}
{{d^3k_2}\over{(2\pi)^3}}k_{12}^2 B(k_1,k_2,k_{12}), 
\end{eqnarray}
\begin{eqnarray}
\nonumber
 S^{(2)} &=& -{9 \over 4}\left< ({\bf\nabla}\delta\cdot{\bf\nabla}\delta)
\nabla^2\delta\right> /\sigma_1^4\\
&=&{9\over {4\sigma_1^4}}\int{{d^3k_1}\over{(2\pi)^3}}
{{d^3k_2}\over{(2\pi)^3}}{\bf k_1\cdot k_2}k_{12}^2 B(k_1,k_2,k_{12}), 
\end{eqnarray}
where $k_{12}=|{\bf k}_1 + {\bf k}_2|$ and $B$ is the bispectrum.
The skewness parameters can be calculated from (Matsubara 2003)
\begin{equation} 
S^{(0)} = (2+E) S_0^{11} - 3S_1^{02}+(1-E) S_2^{11},
\end{equation}
\begin{eqnarray}
\nonumber
S^{(1)} &=& {3\over 2} ({{5+2E}\over 3} S_0^{13}-{{9+E}\over 5}
S_1^{22} \\
&&-S_1^{04}+{{2(2-E)}\over 3}S_2^{13}-{{1-E}\over 5} S_3^{22}),
\end{eqnarray}
\begin{eqnarray}
\nonumber
S^{(2)} &=& 9({{3+2E}\over 15} S_0^{33}-{1\over 5}S_1^{24}-{{3+4E}\over 21}
S_2^{33}\\
&&+{1\over 5}S_3^{24} - {{2(1-E)}\over 35}S_4^{33}), 
\end{eqnarray}
where 
\begin{equation}
E \approx {3 \over 7} \Omega_m^{-1/30}-{{\Omega_{\Lambda}}\over 80} 
(1-{3\over 2}\Omega_{\Lambda} {\rm log_{10}}\Omega_m),
\end{equation}
and
\begin{eqnarray}
\nonumber
S_m^{\alpha \beta} &=& {{\sqrt{2\pi}\over\sigma_0^4}}
({{\sigma_0}\over{\sigma_1 R}})^{\alpha+\beta-2} \int{{\ell_1^2 d\ell_1}\over
{2\pi^2 R^3}}{{\ell_2^2 d\ell_2}\over{2\pi^2 R^3}} P\left({\ell_1\over R}\right) 
P\left({\ell_2\over R}\right) \\
&&\times e^{-\ell_1^2-\ell_2^2}\ell_1^{\alpha-3/2}
\ell_2^{\beta-3/2} I_{m+{1\over 2}}(\ell_1\ell_2). 
\end{eqnarray}
Here $P$ is the linear power spectrum and $I_{\nu}$ are the modified Bessel functions.

We have performed the above integrals to get the skewness parameters 
for our $\Lambda$CDM and SCDM models as a function of smoothing length $R_G$. 
The curves in Figure 8 show the genus-related statistics predicted by 
equation (9) for the $\Lambda$CDM and SCDM models at the present epoch $z=0$. 
In the top panel the RMS density fluctuation $\sigma_0$ and the skewness 
parameters are shown.  The differences between the skewness 
parameters in equation (9) are very small as can be seen here, and 
make the genus as a function of $\nu_f$ rather insensitive to gravitational evolution.
According to the perturbation theory the amplitude of the genus curve remains
the same in the weakly non-linear regime. This is certainly not true 
because the density field starts to build phase correlations and the genus amplitude
starts to decrease as soon as the gravitational evolution occurs.
The behaviours of the $\Delta\nu$, $A_C$, and $A_V$ parameters predicted
by equation (9) are remarkably similar to those seen in Figure 1 and 6.
This is an important confirmation of the Matsubara's theory.
The one major failure of the analytic prediction is for the $A_C$ parameter
at small non-linear scales. At high redshifts $A_C$ decreases below 1 as high density 
clumps merge together as predicted by the analytic theory. But at lower 
redshifts $A_C$ increases to become greater than 1 as the merging rate 
slows down and the speed new clumps appear becomes higher.
\begin{figure}
\begin{minipage}[t]{7cm}
\epsfxsize=8cm
\epsfysize=8cm
\epsffile{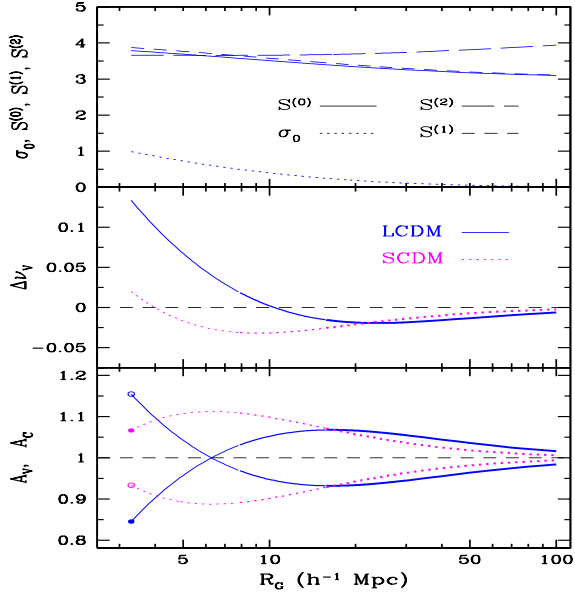}
\end{minipage}
\caption{Genus-related statistics at $z=0$ according to the Matsubara's 
perturbation theory. In the second and third panels thick curves are 
for scales with $\sigma_0 <0.25$, 
medium thickness curves for $0.25\le \sigma_0 <0.5$, and thin curves 
for $0.5\le \sigma_0 <1.0$. In the top panel the RMS deviation of density
fluctuation and skewness parameters are plotted.
\label{fig8}}
\end{figure}

Matsubara (1996) has derived the genus in redshift space in the linear regime. 
The genus in redshift space predicted by the linear theory is
\begin{equation}
G^{(s)} (\nu) = {{3\sqrt{3}}\over 2} \sqrt{C} (1-C) G^{(r)} (\nu), 
\end{equation}
where $G^{(r)}(\nu)$ is the genus curve in real space,
\begin{equation}C = {1\over 3}(1+{6\over 5}{f\over b}+{3\over 7} ({f\over b})^2)/
(1+{2\over 3}{f\over b}+{1\over 5} ({f\over b})^2), 
\end{equation}
$b$ is the bias factor, and 
\begin{equation}
f = {{d {\rm ln}D}\over {d {\rm ln}a}} \approx \Omega_m^{0.6}+{\Omega_{\Lambda}\over 70} (1+{\Omega_m\over 2}), 
\end{equation}
where $D$ is the linear growth factor, and $a$ is the expansion parameter. The formula 
says the redshift space distortion does not alter the shape of the genus curve and 
only affects its amplitude. For matter in our $\Lambda$CDM model $b=0.9$, 
$f=0.468$ and $C=0.414$, and the amplitude drops by a factor of 0.980 while
in the SCDM model which has $b=1.5$, $f=1$, and $C=0.433$ the factor is 0.970.
In the linear regime the genus amplitude calculated 
from redshift space power spectrum is equal to that from real space power 
spectrum since the shape of power spectrum remains the same. This makes the
denominator of the amplitude drop parameter $R_A$ unchanged in redshift space, and
makes $R_A$ itself decrease by above factors in the $\Lambda$CDM and SCDM models.
This can be observed in Figures 1 and 6 at scales $R_G \ge 20 h^{-1}$Mpc.

Matsubara \& Suto (1996) have used N-body simulations to examine the redshift space 
distortion effects on genus. They have reported that the amplitude of the genus 
curve is suppressed more than that expected by linear theory, but that its 
shape in redshift space still remains the same down to weakly non-linear scales.
We have also shown that the redshift space distortion effects on the genus curve
are small in the linear and quasi-linear scales. But it should be pointed out
that the high threshold part of the genus
curve (namely the $A_C$ parameter) is more affected.

Comparisons of numerically calculated genus with theoretical predictions are 
useful. But so far theories have mainly concerned themselves with the 
development of 
non-Gaussianity in the local density field rather than changes in topology. 
Our genus meta-statistics are directly addressing questions about the gravitational 
evolution of topology and the effects of biasing and redshift space 
distortion on topology.

\section{CONCLUSIONS}

We have studied dependence of the genus statistic on cosmology, tracer, 
redshift, and Gaussian smoothing scale. We summarize our findings.

1. The shift parameter $\Delta\nu$ of the genus curve shows a strong scale 
dependence whose characteristics are determined by the shape of the initial 
power spectrum. For the matter distribution in the $\Lambda$CDM model 
the negative-to-positive transition 
occurs at $\sim 9 h^{-1}$ Mpc while it is at about 4 $h^{-1}$ Mpc for SCDM. 
Below these scales gravitation evolution causes strong positive shifts 
starting from high redshifts. At larger scales the gravitational evolution 
produces weak negative shifts.  
This trend enhances as the evolution proceeds.
When the topology tracers are biased peak particles instead of CDM particles, 
evolution of the genus curve is qualitatively similar. But the positive-to-negative
transition scale is different, and at small scales $\Delta\nu$ keeps slowly 
decreasing, which is opposite to the case of CDM matter particles.
We have demonstrated that galaxies can have a topology very different from 
that of matter on small scales. The $\Delta\nu$ parameter measured from the 
dark halo and 
HOD `galaxy' distributions is positive at scales between 5 and 30 $h^{-1}$Mpc,
and becomes more positive due to the combined effects of formation of new 
objects and gravitational clustering. This trend is opposite to that seen
in distributions of matter and biased peak particles at scales larger than
$9 h^{-1}$Mpc. It remains to be seen which value of the shift parameter
the distribution of real galaxies has.
The redshift space distortion systematically decreases $\Delta\nu$ in the
$\Lambda$CDM universe, and the effects are significant at scales $\le 5 h^{-1}$Mpc. 
We can in principle correct for these 
redshift space distortion effects on $\Delta\nu$ by using the observationally 
measured line-of-sight velocity dispersion to correct the shift parameter 
at both large and small scales for the study of primordial density field 
and galaxy formation mechanisms.

2. The amplitude drop parameter is more difficult to measure than the genus
curve alone because it requires additional accurate measurement of 
power spectrum for each tracer of topology. Direct consequence of 
genus amplitude drop is the phase correlation of density field,
and the phase correlation depends on the shape of the initial density 
power spectrum and the biasing.
We have found that the dark halos and the HOD `galaxies' show a very small
genus amplitude drop with $0.9 \le R_A \le 1$ down to the scale $5 h^{-1}$Mpc 
with little scale dependence.
And it hardly changes due to gravitational evolution or peculiar motions.
This might be incompatible with the observationally measured value of about 
0.6 at $5 h^{-1}$Mpc for the IRAS Point Source Catalogue Redshift Survey 
(Canavezes et al. 1988) or for the Center for Astrophysics Survey 
(Vogeley et al. 1994). Satisfying the constraint from the genus amplitude 
drop can be a challenge for HOD modeling of galaxy formation.

3. The void abundance parameter $A_V$ of the dark matter distribution strongly 
increases above the Gaussian value below 10 $h^{-1}$Mpc, but gently drops below 1 
at larger scales as the gravitational evolution proceeds. This is true for 
both $\Lambda$CDM and SCDM models even though their transition scales 
are different. While gravitational evolution always raises $A_V$ at 
small scales, biasing can lower it below 1 as can be seen for the 
biased peak particles, dark halos, and HOD `galaxies'. Even the direction
of evolution of $A_V$ (and $A_C$) can be changed by biasing. 
Since $A_V$ is observed to be less than 1 for the SDSS galaxies (Park et al. 2005), 
this fact is strong evidence for the existence of biasing 
in the distribution of galaxies with respect to matter. 
Fortunately, $A_V$ at small scales is hardly affected by the redshift 
distortion effects, and can give faithful constraints on galaxy formation in
under-dense regions.

We have considered only three mechanisms for galaxy formation. The peak 
biasing scheme is able to explain the stronger clustering amplitude for 
rarer objects (Kaiser 1984). But we have considered only one version 
of the peak biasing scheme where the threshold for galaxies is a step 
function, and have not taken into account merging (cf. Narayanan, Berlind,
\& Weinberg 2000 for various biasing models). The first assumption 
will affect the voids, and the second the clusters. The arbitrariness 
of the threshold function can be removed by using the dark halos which 
have collapsed and are self-bound physical objects. But the dark halos 
also cannot fully trace galaxies because the mass function of dark 
halos continuously extends up to supercluster scales and does not 
show a natural cutoff for galaxies.  Some of them with small masses 
may have lost baryons in high density environments, and have not been 
able to form stars.
The HOD `galaxies' seem to approximate the distribution of real galaxies 
better than the dark halos because the $A_V$ parameter is closer to 
observations for the HOD `galaxies' even though the cluster abundance 
parameter $A_C$ may be still too small (cf. Park et al. 2005).

While galaxy formation and evolution in a high density environment are
complicated due to the merging of structures and the high ram pressure 
of the intergalactic medium (Gunn \& Gott 1972), 
it may be relatively easier to model galaxy formation in low 
density regions.
For example, biased peak particles, dark halos, and HOD `galaxies' all
produce $A_V$ smaller than 1 consistent with observations. But their
$A_C$ are wildly different. Therefore,
even though identification of the biased peaks and dark halos with galaxies 
is not a good approximation in high density regions,
they seem to trace galaxies better 
in under-dense regions for which there is a sensitive measure 
like the void abundance parameter $A_V$. It should be emphasized that the 
cosmic voids have not been studied as extensively as clusters and 
superclusters, and it is only in recent years that voids have begun 
to be actively studied 
(Hoyle \& Vogeley 2004; Sheth \& van de Weygaert 2004; Rojas et al. 2004;
Colberg et al. 2004).
After all, voids are not the 
places where there is nothing, but places where history of the 
universe is better kept. With the advent of new large observational 
data sets like the SDSS sample we can hope better to understand the 
formation mechanism of each species of galaxies through the topology 
analysis. We are currently studying the topology of galaxies divided 
into subgroups with different internal physical properties.

\acknowledgments


CBP is supported by the Korea Science and Engineering Foundation
(KOSEF) through the Astrophysical Research Center for the Structure and
Evolution of the Cosmos (ARCSEC) and through the grant R01-2004-000-10520-0.
CBP and JHK also acknowledge the support of Korea Institute of Science and 
Technology Information through the `Grand Challenge Support Program'.
JRG is supported by NSF grant AST04-06713.

\clearpage

\clearpage




\end{document}